\documentclass[prb,aps,amsfonts,amssymb,floatfix,showpacs,twocolumn,superscriptaddress]{revtex4}
\usepackage{graphicx}
\usepackage{dcolumn}
\usepackage{bm}
\newcommand{\kvec}{{\bf k}}
\newcommand{\qvec}{{\bf q}}

\begin{document} 
\title{Gutzwiller Magnetic Phase Diagram of the Cuprates}
\author{R.S. Markiewicz}
\affiliation{ Physics Department, Northeastern University, Boston MA 
02115, USA}
\affiliation{SMC-INFM-CNR and Dipartimento di Fisica, Universit\`a di Roma
``La Sapienza'', P. Aldo Moro 2, 00185 Roma, Italy}
\affiliation{ISC-CNR, Via dei Taurini 19, I-00185 Roma, Italy}
\author{J. Lorenzana}
\affiliation{SMC-INFM-CNR and Dipartimento di Fisica, Universit\`a di Roma
``La Sapienza'', P. Aldo Moro 2, 00185 Roma, Italy}
\affiliation{ISC-CNR, Via dei Taurini 19, I-00185 Roma, Italy}
\author{G. Seibold}
\affiliation{Institut F\"ur Physik, BTU Cottbus, PBox 101344, 03013 Cottbus, 
Germany}
\author{A. Bansil}
\affiliation{ Physics Department, Northeastern University, Boston 
MA 02115, USA}
\begin{abstract}
A general constructive procedure is presented for analyzing magnetic instabilities in 
two-dimensional materials, in terms of [predominantly] double nesting, and applied to 
Hartree-Fock HF+RPA and Gutzwiller approximation GA+RPA calculations of the Hubbard model.
Applied to the cuprates, it is found that competing magnetic interactions are present only 
for hole doping, between half filling and the Van Hove singularity.  While HF+RPA 
instabilities are present at all dopings (for sufficiently large Hubbard $U$), in a 
Gutzwiller approximation they are restricted to a doping range close to the range of 
relevance for the physical cuprates.  The same model would hold for charge instabilities, 
except that the interaction is more likely to be $q$-dependent.
\end{abstract} 
\maketitle


\section{Introduction}

The charge and magnetic instabilities of one dimensional materials are well understood 
in terms of Fermi surface (FS) nesting.  Here it is shown that for two-dimensional 
materials, features in maps of the bare susceptibility can be understood in terms of a 
series of FS `nesting curves', and the dominant instabilities are generally related 
to {\it double nesting} features.  The analysis shows how to locate these nesting 
features in momentum (${\bf q}$) space, often providing analytical expressions.  As an 
application, the full evolution with doping of the leading magnetic instabilities for 
several families of cuprates is presented, both in the conventional Hartree-Fock (HF) plus 
RPA (HF+RPA) and in a Gutzwiller approximation (GA+RPA) calculation.  The analysis 
provides a pseudogap candidate and makes the surprising prediction that the 
`checkerboards' seen in scanning tunneling microscopy (STM) studies are {\it not} the same 
phase as the `stripes' in La$_{2-x}$Sr$_x$CuO$_4$ (LSCO).  

The present results should find extensive utilization.  First, for weakly correlated 
two-dimensional systems the HF+RPA results provide an essentially complete solution to the 
nesting problem, as long as the interaction $U$ is ${\bf q}$-independent.  The GA+RPA 
extends the results into the intermediate coupling regime.  In both cases, the magnetic 
instabilities are determined by zeroes of the Stoner denominator, 
\begin{equation}
1-U_{eff} \chi_0({\bf q},\omega =0).
\label{eq:G1}
\end{equation}
Here for cuprates $U_{eff}$ is the Hubbard $U$ in HF+RPA, and a more 
complicated ${\bf q}$-dependent object $U_{GA}({\bf q})$ in the GA+RPA calculation.
Thus, the leading HF+RPA instability is simply associated with the maximum of the 
bare susceptibility $\chi_{0M}=\max_{\bf q}\chi_0({\bf q},0)$, while the leading Gutzwiller 
instability can be shifted by the ${\bf q}$-dependence of $U_{GA}({\bf q})$.  It will be clear 
that the same analysis can be extended to any two-dimensional material.  

The cuprates appear to be in an intermediate coupling regime where the Gutzwiller 
results can be expected to provide a good approximation to the phase diagram 
at $T=0$.   Thus, in the cuprates, recent quantum Monte Carlo (QMC)\cite{MJM} and 
`quasiparticle-GW' (QP-GW)\cite{water3} calculations have been able to reproduce 
experimental ARPES spectra of optimally and overdoped cuprates, starting essentially 
from LDA bands and calculating the self-energy self consistently.  In the QP-GW 
approach, the self energy is calculated as a convolution over a renormalized RPA 
susceptibility.  Not only is the low-energy dispersion renormalization reproduced, but 
also the `waterfall' effect\cite{Ale}, which represents the dressing of low-energy 
quasiparticles by (mainly) spin fluctuations.  Extension of these results to the 
hole-underdoped regime will require identification of the phase or phases responsible 
for the pseudogap.  Since these are most likely to be incommensurate density wave or 
`stripe' phases, the QMC calculations have a severe problem of limited
${\bf q}$-resolution,  
while the QP-GW calculations are ideally suited to handle this.  Such an approach has 
had considerable success with electron-doped cuprates, where the leading instability is 
commensurate $(\pi ,\pi)$. (We set the lattice constant $a\equiv 1$.)

The paper is organized around the Stoner criterion as follows.  Sections 2-3 describe 
the calculations of the zero frequency bare susceptibility $\chi_0({\bf q})$, and 
determining its maximum $\chi_{0M}$.  In Section 2 we introduce the concept of 
nesting curve, associated with the nesting criterion ${\bf q}=2{\bf k}_F$, where ${\bf k}_F$ is the 
(anisotropic) Fermi wave vector, and we demonstrate that $\chi_{0M}$ is generally 
associated with a double nesting.  We apply this concept to the cuprates, taking 
the hopping parameters from tight-binding one-band fits to the LDA dispersions, and 
show that over most of the doping range there is a unique $\chi_{0M}$, with a 
surprising electron-hole symmetry.  In Section 3 we extend this analysis to the more 
complicated hole doping regime for doping $x$ between half filled and the doping of the Van 
Hove singularity (VHS).  Here competing magnetic phases are found, and the dominant phase is 
sensitive to material parameters, being different for different cuprates.  Section 4 
describes the GA+RPA technique and introduces the corresponding $U_{eff}=U_{GA}({\bf q})$.  
Section 5 presents the resulting Gutzwiller magnetic phase diagrams.  It is found that the 
phase diagram of La$_{2-x}$Sr$_x$CuO$_4$ (LSCO) is distinct from that of most other cuprates, 
but that for all cuprates, using a bare Hubbard $U=8t$, the paramagnetic state is unstable at 
the GA level for all relevant dopings, including the overdoped regime. A discussion is 
presented in Section 6, and conclusions in Section 7.

\section{Susceptibility Plateaus}

The present calculations are based on tight-binding parametrizations of typical 
dispersions for single-layer cuprates, including models of La$_{2-x}$Sr$_x$CuO$_4$ 
(LSCO), Nd$_{2-x}$Ce$_x$CuO$_4$ (NCCO), and Bi$_2$Sr$_2$Cu$_1$O$_{6}$ (Bi2201), Table I 
in Appendix~\ref{sec:band-parameters}.  For NCCO, Bi-2201(2), and LSCO(2), the parameters are based on a 
tight-binding fit to the LDA bands\cite{Arun3}, while sets Bi-2201(1)
and LSCO(1) are direct  
fits to the experimental bands.  In all cases, $k_z$ dispersion is
neglected, approximating  
the cuprates as two-dimensional.  

\begin{figure}
\leavevmode  
\resizebox{8.2cm}{!}{\includegraphics{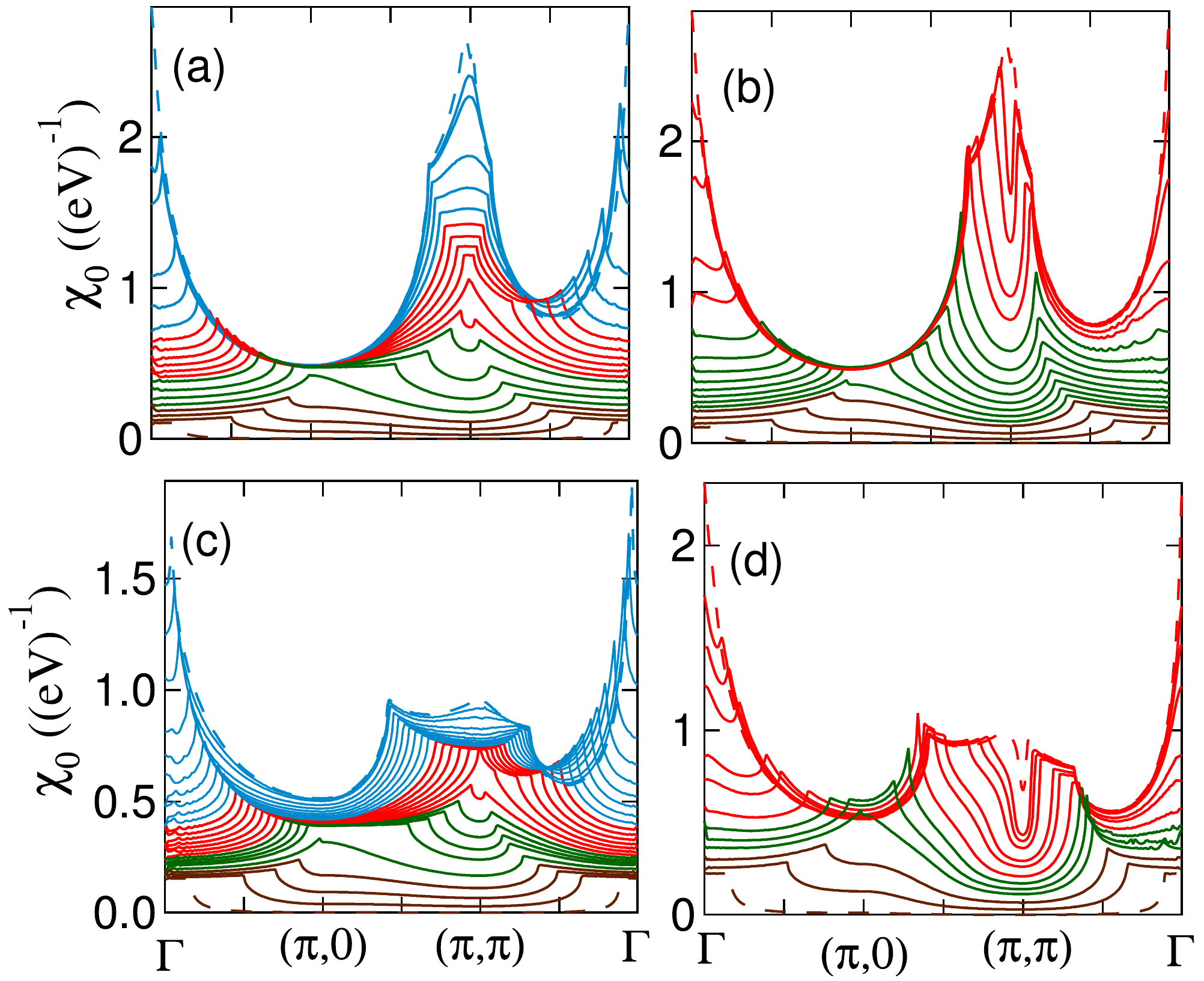}}
\vskip0.5cm
\caption{(Color online.)
(a) Susceptibility $\chi'_0(\omega =0)$ for LSCO as a function of ${\bf q}$ for a series of
dopings from $x_{VHS}=0.207$ (blue dashed line) to $x=-0.99$ (brown dashed line).  [Blue curves are for 
hole doping, $x>0$, others for electron doping, $x\le 0$.]  Note the evolution from a plateau 
near $(\pi ,\pi )$ (blue and red lines) to one near $(\pi ,0)$ (green) to one near $\Gamma$ 
(brown).  Band parameters appropriate to LSCO(2) (Table I).
(b) Same as (a), except for hole dopings $x=0.207$ (red dashed line) to $x=0.99$ (brown 
dashed line).  Note similar evolution of plateaus.
(c,d) Similar to (a,b), except for Bi-2201 (2 in Table I).  Note overall similarity, 
except for curvature near $(\pi ,\pi ).$  Dashed curves correspond to $x$ = (c): 
-0.99 (brown) or 0.43 (blue); (d): 0.44 (red) or 0.99 (brown). 
Curves are generally spaced by $\Delta x=0.05$, except for (1) higher density near 
points of rapid change (e.g., the VHS), (2) $\Delta x=0.1$ near top and bottom of band, 
and (3) end points at $x=\pm 0.99$. 
}
\label{fig:1}
\end{figure}  

The real part of the susceptibility $\chi_0'({\bf q},\omega =0)$ is dominated by a
series of plateaus, with the largest susceptibility systematically shifting from one
plateau to another as a function of doping.  Figure~\ref{fig:1} shows that these and 
related features dominate $\chi_0'$ over the full doping range, and further reveals a 
striking quasi-electron-hole symmetry of the evolution.  However, instead of being 
symmetric about $x=0$, the evolution is symmetric about the doping of the Van Hove 
singularity (VHS), $x_{VHS}$, where the Fermi energy coincides with the VHS, $E_F=E_{VHS}$, 
and the density of states (DOS) diverges logarithmically.  Figure~\ref{fig:1}(a) shows 
$\chi_0'$ calculated along the high symmetry axes of LSCO, for the full electron-doping 
range $n=1-x$ from 1 to 2, with an extension to the hole doping of the VHS (blue curves).  
Near half filling the susceptibility is dominated by the well-known plateau\cite{AAA} near 
$(\pi ,\pi )$ (red and blue curves), which peaks at the VHS.  As electron-doping increases, 
the peak shifts to a second plateau near $(\pi ,0)$ (green curves), then to a third near 
$\Gamma$ for a nearly full band (brown curves).  The {\it same} sequence is followed for 
hole doping, Fig.~\ref{fig:1}(b). 

In addition to these plateaus, there is an additional $\Gamma$-centered feature, which is 
prominent near the doping of the VHS, but has largely disappeared by half filling.  This 
feature will be referred to as the antinodal nesting (ANN) plateau.  Bi2201 displays the same 
plateaus, with the same quasi-electron-hole symmetry, Fig.~\ref{fig:1}(c,d).  
A key difference is that the $(\pi ,\pi )$-plateau of Bi2201 is convex over a considerably 
wider doping range than in LSCO, before turning concave near the VHS.  This has a consequence 
that the ANN peak dominates over a wide doping range in Bi2201, but only in the immediate 
vicinity of the VHS in LSCO.  At the VHS, the bare susceptibility diverges logarithmicly 
both at $\Gamma$ (where it is equal to the DOS) and at $(\pi ,\pi )$, but the latter 
divergence is quite weak and not apparent in the numerical calculations of Fig.~1. 
Despite the apparent complexity of these evolving susceptibility patterns, the plateau 
evolution can be understood in detail, with analytic formulas for the positions of all 
dominant peaks.

\begin{figure}
\leavevmode  
\includegraphics[width=9cm,clip=true]{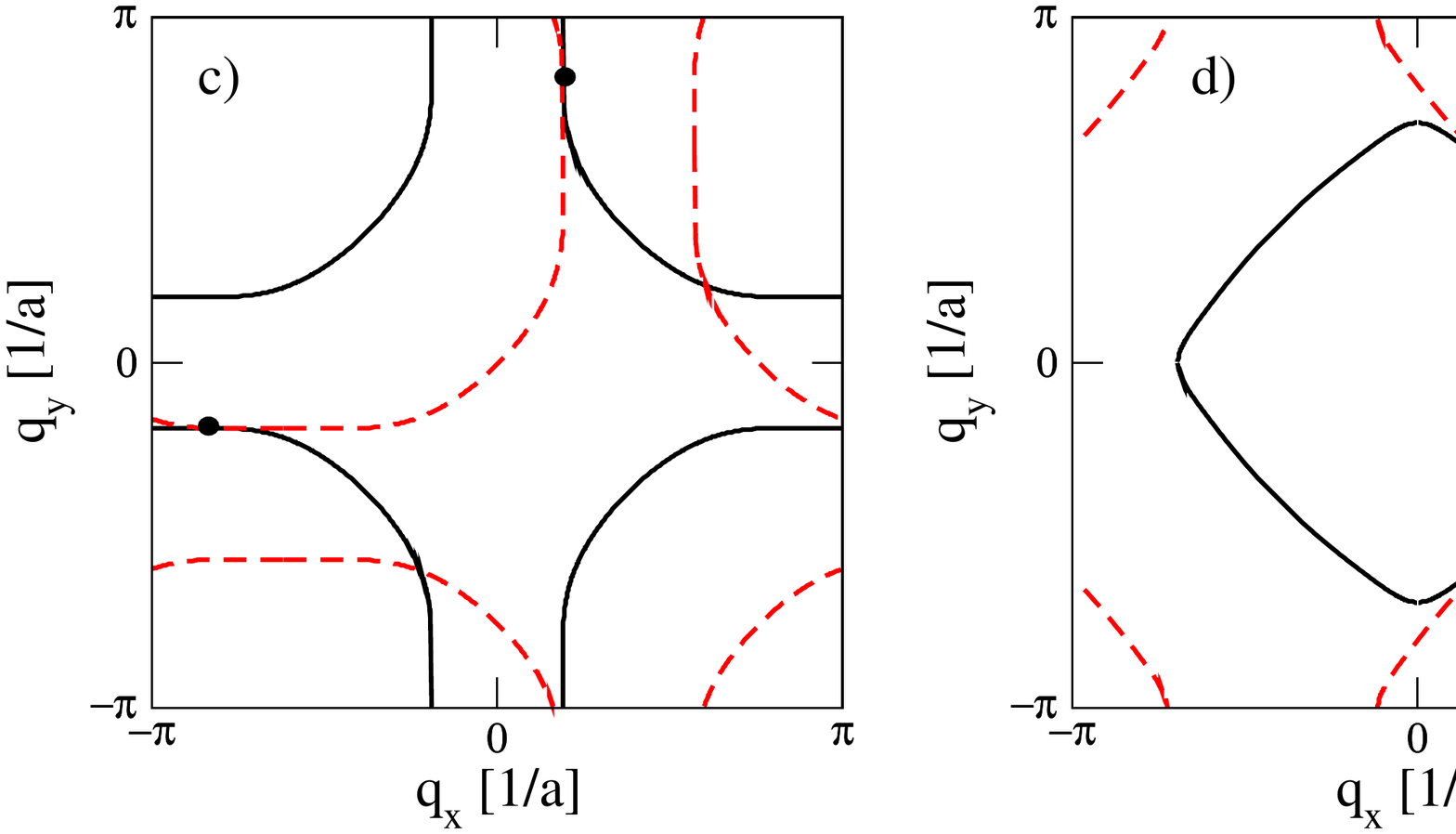}
\includegraphics[width=9cm,clip=true]{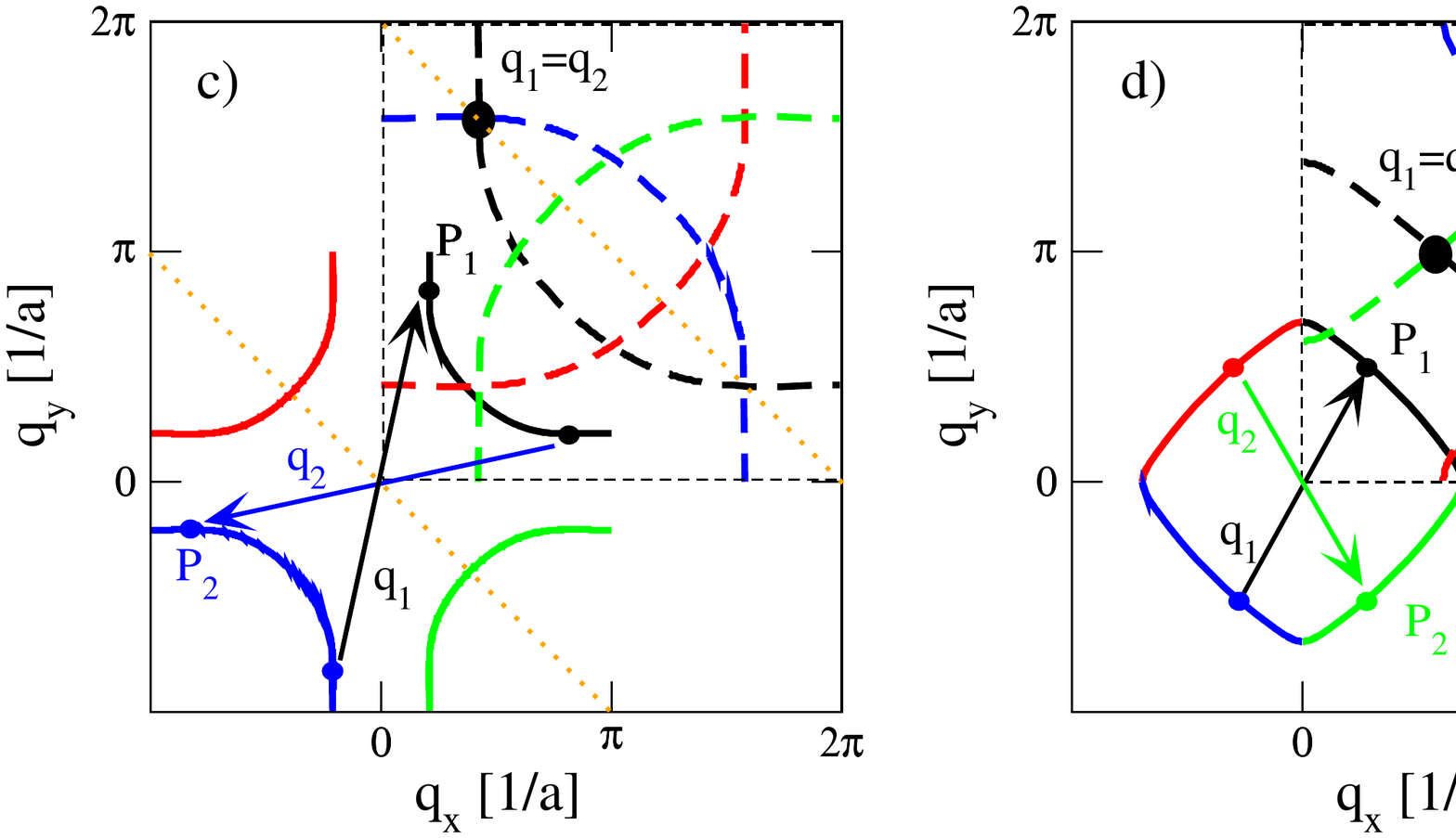}
\vskip0.5cm
\caption{(Color online.)(a) Hole-like Fermi surface (solid) shifted by the double nesting vector 
${\bf q}= (0.38,1.62)\pi/a$ and
folded back into the first BZ (dashed). 
(b) Electron-like Fermi surface (solid) shifted by the double nesting vector 
${\bf q}= (0.58,1)\pi/a$ and folded back into the first BZ (dashed). 
The (double) nesting points are marked 
by a solid dot.
(c,d) Construction of nesting curves (dashed) for hole-like (c) and electron-like (d) FS's (solid lines)
from condition Eq. \ref{eq:G2}. The FS is shown in the first BZ whereas 
the nesting curves are folded into the momentum space $0\le q_{x/y} \le 2\pi$
as defined by the dashed square. The arrows indicate scattering processes which
lead to `double nesting' as explained in the text. The dotted lines corresponds
to the boundary of the magnetic BZ in the first and enlarged zone. 
(a,c) Parameters for NCCO, $x=-0.15$;
(b,d) Parameters for LSCO(1), $x=0.41$.}
\label{fig:new}
\end{figure}  

\begin{figure}
\leavevmode  
\resizebox{8.2cm}{!}{\includegraphics{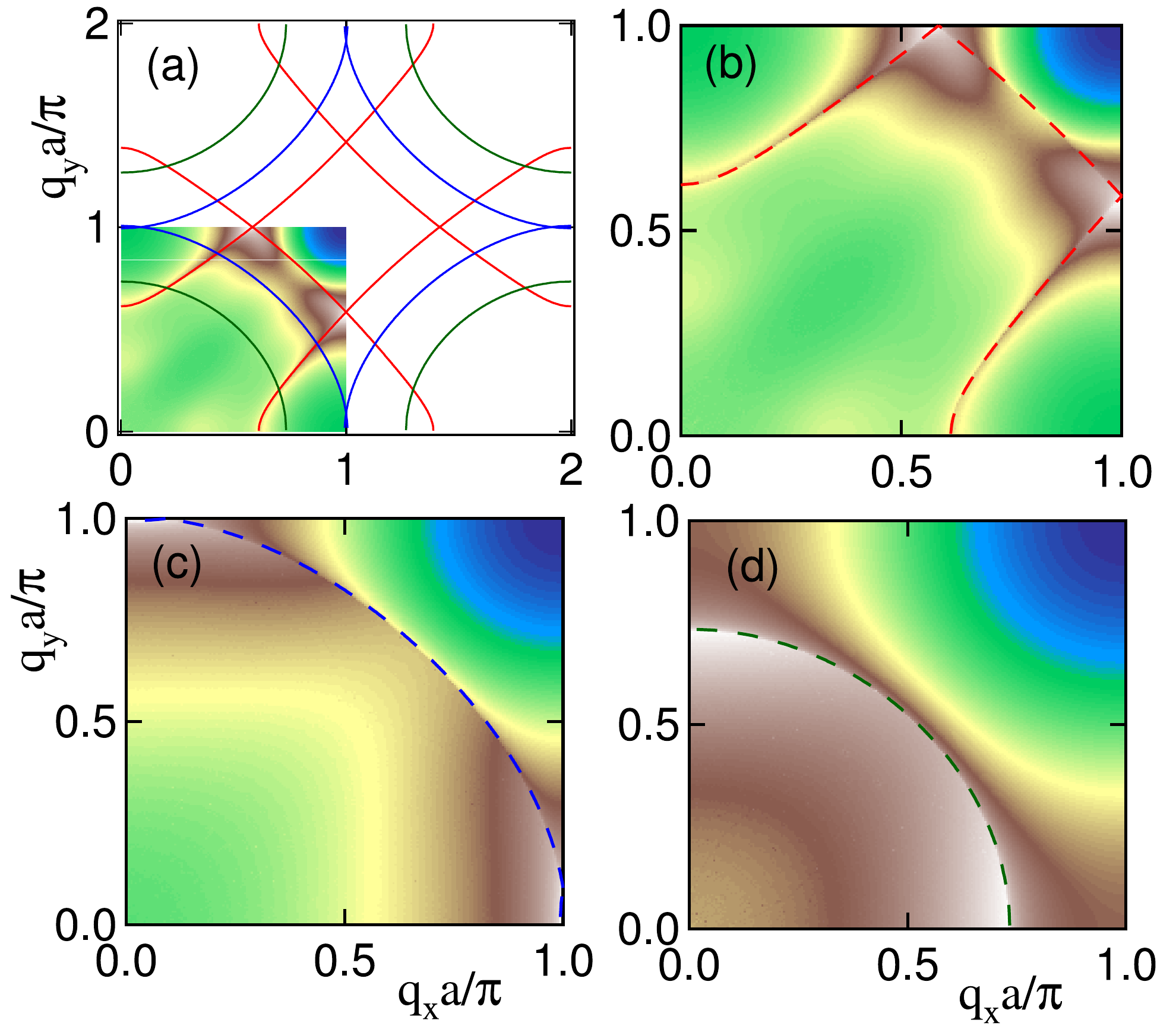}}
\vskip0.5cm
\caption{(Color online.)
(a) Nesting maps $q=2k_F$ for LSCO, dispersion 1, and $x$ = 0.41 (red
curves), 0.62 (blue), and 0.79 (green). Corresponding susceptibility $\chi'_0(\omega
=0)$ as a function of $q$ for a series of hole dopings from $E_F$ [$x$] = -0.16 [0.41]
(b), -0.3 [0.62]] (c), and -0.5~eV [0.79] (d).  In all figures, whites are largest
$\chi$s, blues are smallest.
}
\label{fig:2}
\end{figure}

The generic evolution of the plateaus from $(\pi ,\pi )\rightarrow(\pi ,0 
)\rightarrow(0,0)$ can be understood with reference to the `nesting curves', Fig.~\ref{fig:new}. 
For the generic case of two Fermi surface segments, a nesting curve can be defined as the 
locus of all points ${\bf q}={\bf k_{F1}}-{\bf k_{F2}}$, where ${\bf k_{Fi}}$ is a 
point on the $i$th FS, FS$_i$, with the restriction that when FS$_1$ is shifted by 
${\bf q}$ it is tangent to FS$_2$.  For the cuprates, there is usually just a single FS 
section [an exception is given below in Section III, Fig.~6], and the nesting curves 
simplify to plots of
\begin{equation}
{\bf q}=2{\bf k}_F, 
\label{eq:G2}
\end{equation}
for any Fermi momentum ${\bf k}_F$, Fig.~\ref{fig:new}.  Since the point $(\pi ,\pi )$ lies on 
horizontal and vertical planes of reflection symmetry for the susceptibility, the nesting curves 
must be supplemented by their reflections about the lines $q_x=\pi$, $q_y=\pi$.  
Fig.~\ref{fig:2} shows three sets 
of nesting curves corresponding to three hole dopings for LSCO (dispersion 1), $x$  = 0.41 (red 
curves), 0.62 (blue), and 0.79 
(green).  Frames (b-d) show the corresponding susceptibility maps $\chi_0'({\bf q})$, and it 
can be seen that the ridges in $\chi_0'$ are exactly given by the nesting curves. [For 
convenience, the $x=0.41$ data is replotted in Fig.~3(a).  Note that the susceptibility maps 
are plotted over the range of $q_x,~q_y$ between 0 and $\pi$, whereas the nesting curves are 
plotted over the wider range 0 to $2\pi$.]  Furthermore, the dominant peaks in $\chi_0'$ 
correspond to the intersection of two nesting curves.  By drawing the original and 
${\bf q}$-shifted Fermi surfaces (FSs), Figs.~\ref{fig:new}(a),~(b), 
it can be seen that this overlap corresponds to the simultaneous nesting of two different 
sections of FS. Hence the term `double nesting'.

As shown in Figs.~\ref{fig:new}(c),~(d) this kind of `double nesting' can originate from either
the scattering between points on opposite segments (cf. example in Fig.~\ref{fig:new}(c)) or
adjacent segments (cf. example in Fig.~\ref{fig:new}(d)) of the FS.
In the example shown in Fig.~\ref{fig:new}(c) we denote the scattered states on the FS as 
${\bf P_1}=(\delta_\perp, \pi-\delta_\parallel)$ and ${\bf P_2}=(-\pi+\delta_\parallel, 
-\delta_\perp)$ which yield the points on the nesting curves ${\bf q_1}=2 {\bf 
P_1}=(2\delta_\perp, 2\pi-2\delta_\parallel)\equiv (2\delta_\perp, -2\delta_\parallel)$ and 
${\bf q_2}=2 {\bf P_2}=(-2\pi+2\delta_\parallel, -2\delta_\perp) \equiv (2\delta_\parallel, 
-2\delta_\perp)$. Thus in this case `double nesting' (${\bf q_1}={\bf q_2}$)
occurs when $\delta_\perp = \delta_\parallel$ which generally can only be fulfilled for points
near the antinodes of hole-like FS's.  These are thus referred to as antinodal nesting (ANN) 
features.  More common is the situation sketched in Fig.~\ref{fig:new}(d)
where we consider the scattered states ${\bf P_1}=(\delta_\perp, \pi-\delta_\parallel)$
and ${\bf P_2}=(\delta_\perp, -\pi+\delta_\parallel)$. In this case the `double nesting' 
condition ${\bf q_1}={\bf q_2}$ can only be fulfilled for $\delta_\parallel=0$ (which is trivial
since initial and final states of the two scattering processes are identical)  or 
$\delta_\parallel=\pi/2$.  The latter condition implies that this kind of `double nesting' 
generally affects the nodal states and leads to scattering vectors ${\bf q}$ close to 
$(\pi/a,\pi/a)$, generating the $(\pi ,\pi )$-plateau.
 
A slightly different point of view might help clarify the role of double nesting.  
Tangency of two Fermi surface segments [`nesting'] means a stability of FS overlap,
in that the nesting persists if one displaces the nesting vector in a particular
direction.  Then double nesting means tangency along two surface segments, so nesting 
persists if one displaces the nesting vector in two (rather than one) directions.

\begin{figure}
\leavevmode  
\resizebox{8.2cm}{!}{\includegraphics{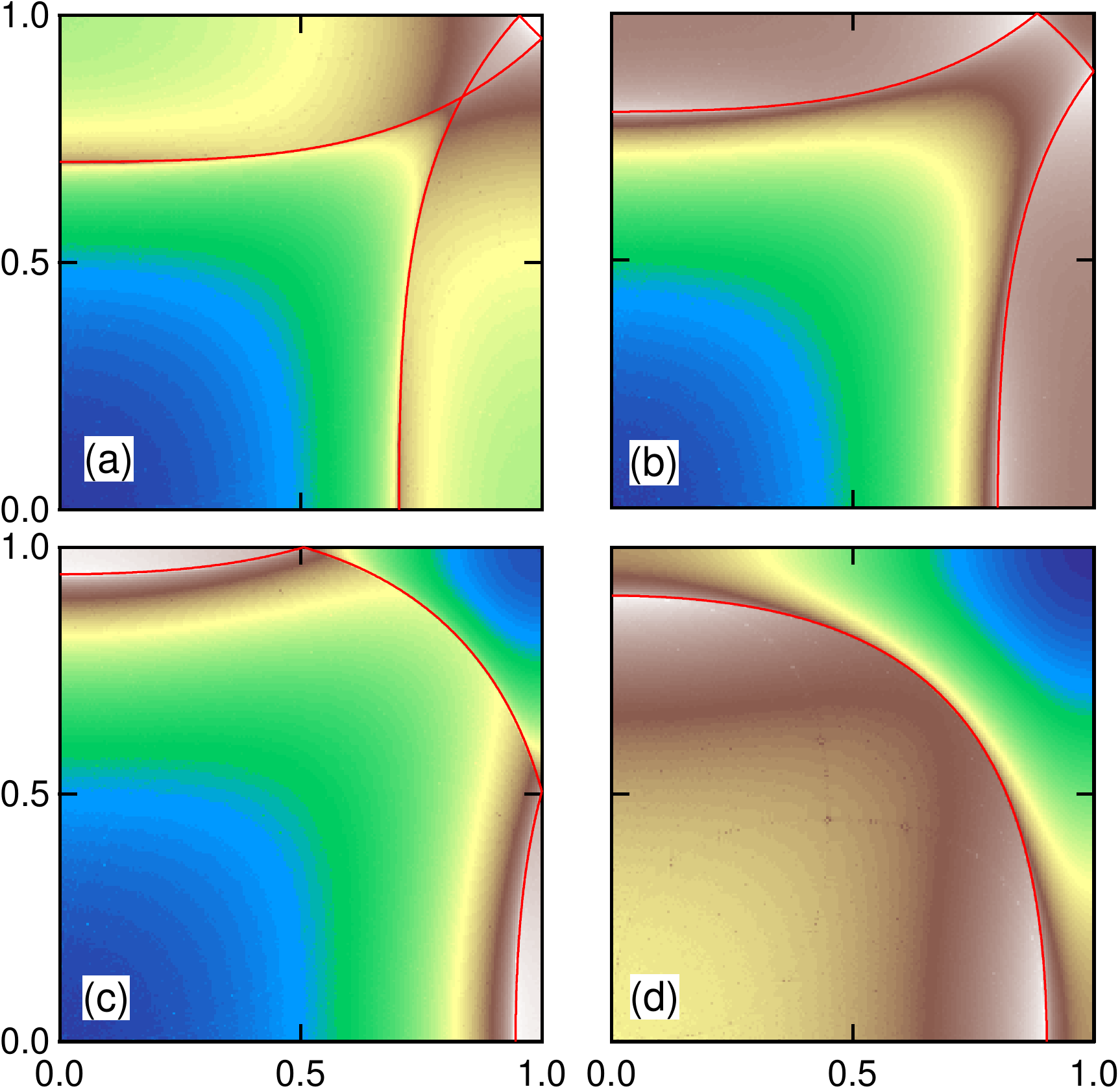}}
\vskip0.5cm
\caption{(Color online.)
Susceptibility $\chi'_0(\omega =0)$ as a function of $q$ with superposed
nesting maps $q=2k_F$ for NCCO, for a series of electron dopings ($x<0$) from
$x$ [$E_F$] = -0.26 [0.10] (a), -0.37 [0.20] (b), -0.52 [0.35] (c), and
-0.65 [0.50~eV] (d).
}
\label{fig:3}
\end{figure}

The origin of the plateau transitions is now apparent: as hole doping increases, the FS shrinks to a 
small pocket near $\Gamma$ before disappearing.  The nesting curves shrink in a similar fashion, 
but with a doubled radius, since ${\bf q}=2{\bf k}_F$.  Thus the dominant overlap shifts from near $(\pi 
,\pi )$ at the VHS (Fig.~2(b)) towards $(\pi ,0)$ (Fig.~2(c)), and finally towards $\Gamma$, 
in Fig.~2(d), thereby explaining the plateau evolution.  For electron doping, the Fermi 
surface ultimately closes at $(\pi ,\pi )$, leading to the same sequence of plateau 
transitions, as illustrated in Fig.~\ref{fig:3}, where the nesting maps are directly 
superposed on the susceptibility curves.  In the doping range relevant 
to the cuprates, the physics of NCCO is dominated by the $(\pi ,\pi )$ plateau, 
Fig.~\ref{fig:3}a and red curves in Fig.~1(a), which shrinks to a point at the end 
of the `hot-spot' regime.  


In all cases, the dominant peak lies along a high symmetry axis, and the doping dependence 
of its position can readily be found from the dispersion
\begin{eqnarray}
     E({\bf k})=-2t[c_x(a)+c_y(a)]-4t'c_x(a)c_y(a) \nonumber \\
     -2t''[c_x(2a)+c_y(2a)] \nonumber \\
     -4t'''[c_x(2a)c_y(a)+c_y(2a)c_x(a)] \> , 
\label{eq:2}
\end{eqnarray}
where
\begin{equation}
     c_i(\alpha a)\equiv \cos(\alpha k_ia) \>,
\label{eq:0d}
\end{equation}
and $\alpha$ is an integer (or half-integer).  For this dispersion, the VHS is generally at 
$(\pi ,0)$, or $E_{VHS}=4t'-2t''$.  The positions of the peaks in 
Fig.~\ref{fig:2}(b,c) and Fig.~\ref{fig:3}(b,c) lie along the zone boundary 
($q_xa=\pi$) with  
\begin{equation}
q_ya =2\arccos[\pm\sqrt{b^2-c}-b],
\label{eq:1x}
\end{equation}
with $b=(t-2t''')/[4t'']$ and $c=[E_F-4t'']/[4t'']$.  For 
electron [hole] doping the peak is exactly at $(\pi ,0)$ when $E_F=+[-]2(t-2t''')$.  
Beyond this point, the peak is at $q_y =0$, $q_x$ given by Eq.~\ref{eq:1x} with
$b=(t-[+]2t'+2t''')/[4(t''-[+]2t''')]$ and $c=[E_F-[+]2(t-2t''')]/[4(t''-[+]2t''')]$.
Since a peak at $\Gamma$ corresponds to ferromagnetism, the above reproduces the common 
finding that a nearly empty or nearly full band tends to be ferromagnetic.  
The ANN peak ${\bf q_1}={\bf q_2}$ in Fig.~\ref{fig:new}a satisfies 
$cos(q_ya)=[E_F-2t']/[2(t'-2t'')]$.  Thus the peak 
susceptibility is generally associated with double nesting.  The only exception we have found 
to this is a tendency to remain commensurate for a finite doping range about high symmetry 
points such as $(\pi ,\pi)$ or $\Gamma=(0,0)$,\cite{foot1} -- a form of Van Hove 
nesting\cite{RiSc}.

\section{Competing Phases and Ferromagnetism}

\begin{figure}
\leavevmode  
\resizebox{8.2cm}{!}{\includegraphics{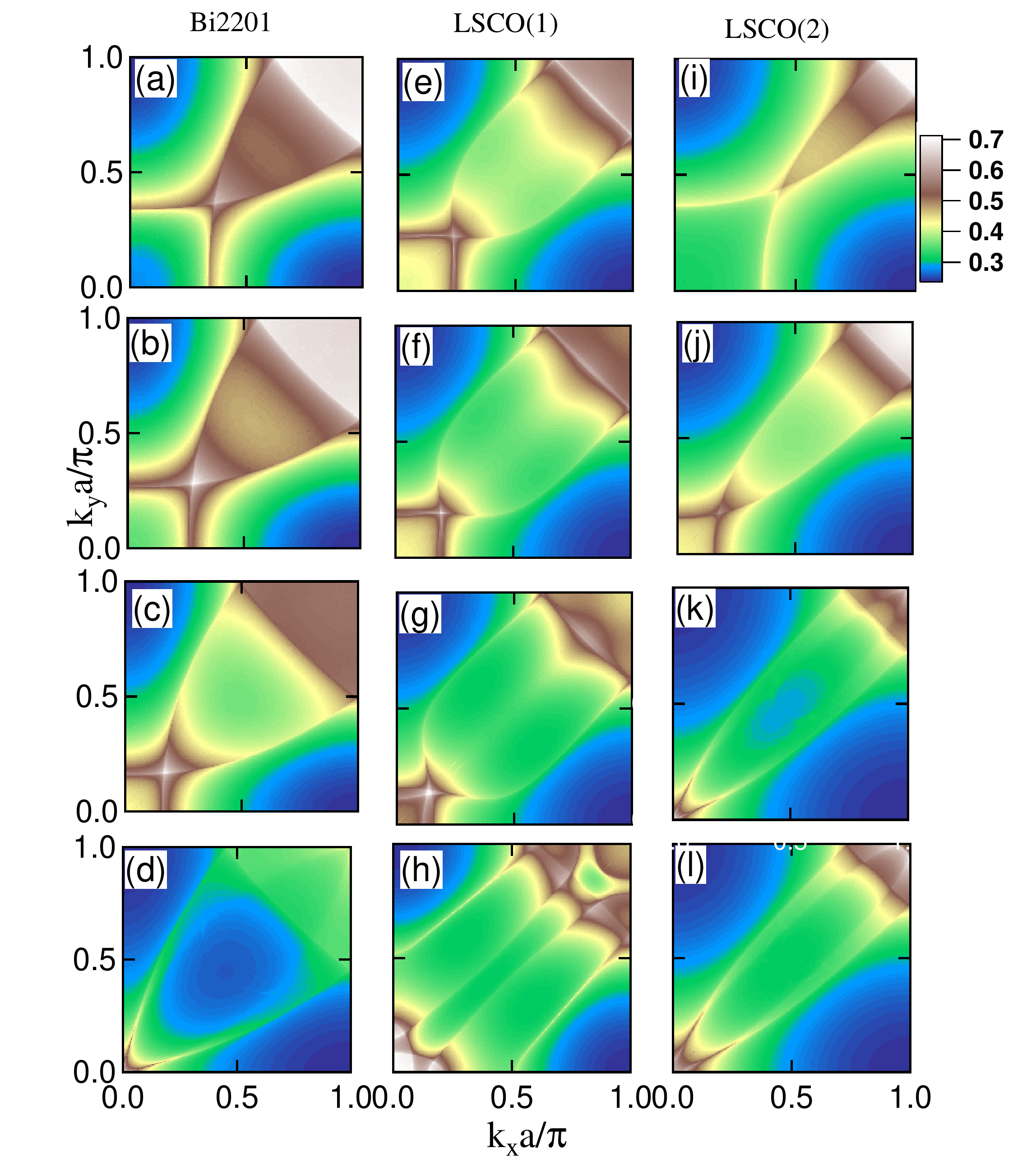}}
\vskip0.5cm
\caption{(Color online.)
Susceptibility maps for Bi2201 (a-d), LSCO(1) (e-h), and LSCO(2) (i-l) at a series of
increasing dopings.  $x$ [$E_F$] = 0.12 [-0.20] (a), 0.20 [-0.25] (b), 0.31 [-0.30]
(c), 0.40$\simeq x_{VHS}$ [0.33] (d); 0.19 [-0.12] (e), 0.23 [-0.13] (f), 0.29 [-0.14]
(g), ($x_{VHS}\simeq 0.33$), 0.37 [-0.15] (h); 0.0 [-0.05] (i), 0.14 [-0.10] (j),
0.20$\simeq x_{VHS}$ [-0.111] (k), and 0.21 [-0.1112 eV] (l).
}
\label{fig:4}
\end{figure}  

While the above sequence is completely generic, holding for all cuprates and being 
electron-hole symmetric, the additional features associated with the VHS are 
considerably more variable.  Fig.~\ref{fig:4} illustrates the low-hole doping regime for 
three dispersions proposed for the cuprates.  At half filling, all three have a peak 
susceptibility on the $(\pi ,\pi )$ plateau, associated with conventional staggered 
antiferromagnetism (SAF).  At low doping the peak is either commensurate or $(\pi ,\pi 
-\delta)$ incommensurate, with $\delta$ increasing with doping.  For finite hole doping 
a new feature emerges, a roughly $+$-shaped peak along the zone diagonal, at $(\pi 
-\delta ,\pi -\delta)$.  From the nesting curves, it can be seen that the peak is 
associated with nesting of the flat sections of the bands near $(\pi ,0)$ -- hence the 
name antinodal nesting (ANN) -- but the largest susceptibility lies along the zone 
diagonal, where both $(\pi ,0)$ and $(0,\pi )$ nesting occur simultaneously, Fig.~\ref{fig:new}a.  
This ANN feature has an interesting relation with `hot spot' physics\cite{hotspots}.  
A `hot spot' is a point of the Fermi surface which is simultaneously on 
the antiferromagnetic (AF) zone boundary [diagonal of the paramagnetic Brillouin 
zone].  
From Fig.~\ref{fig:new}(c) 
it becomes apparent (cf. large dot) that the image of the AF zone boundary in the extended ${\bf 
q}$-map intersects the `nesting curves' exactly at the points of `double nesting'.  But the 
image of the AF zone boundary gets folded in the ${\bf q}$-map onto the zone diagonal, $\Gamma 
\rightarrow (\pi ,\pi )$, thereby generating the usual ANN feature.
 Thus all the diagonal ANN peaks in ${\bf q}$ arise from hot 
spots in $k$.  As the doping of the VHS is approached the hot spots move toward $(\pi 
,0)$ and the nesting curve moves to $\Gamma$.  Since the susceptibility at $\Gamma$ is 
equal to the DOS, it diverges at the VHS, thereby controlling the magnetic 
instability.  

The appearence of the ANN peak leads to a competition between two different kinds of 
magnetic order, and the doping evolution of the susceptibility maps 
diverges.  For most dispersions studied, including the left and central columns of 
Fig.~\ref{fig:4}, the ANN intensity grows and the dominant peak changes discontinuously 
from the SAF plateau to ANN nesting. 
For the Bi-2201 dispersion (left column) this ANN peak becomes dominant at about 
$x$=0.2, and evolves smoothly to $\Gamma$ at $x$=0.4.  If one unfolds 
the nesting curve, one sees that this happens exactly at the doping of the VHS, when the 
FS passes through $(\pi ,0)$.  The central column, corresponding to an extreme 
dispersion proposed for LSCO to enhance antinodal nesting, displays a much more 
complicated nesting map over a limited doping range close to the VHS, but also displays 
a dominant peak at $\Gamma$ exactly at the VHS.  This map will be explained below.  
Finally, for the right column, corresponding to a more conventional dispersion for LSCO, 
the ANN peak is weaker than the SAF peak except in the immediate vicinity of the VHS.  
In this case, the dominant susceptibility peak remains commensurate at $(\pi ,\pi )$ 
until the doping of the VHS, jumps to $\Gamma$ at the VHS, then jumps back and smoothly 
evolves to $(\pi ,\pi -\delta)$ incommensurate [this sequence can also be followed in 
Fig.~1(a),(b)].  Note further, in Fig.~5(i,j), that there is a wide doping range where the 
susceptibility peak remains commensurate, at $(\pi ,\pi )$.  [Since $(\pi ,\pi )$ lies 
along several mirror planes, the corresponding susceptibility is in general a local 
maximum or minimum.]

\begin{figure}
\leavevmode  
\resizebox{8.2cm}{!}{\includegraphics{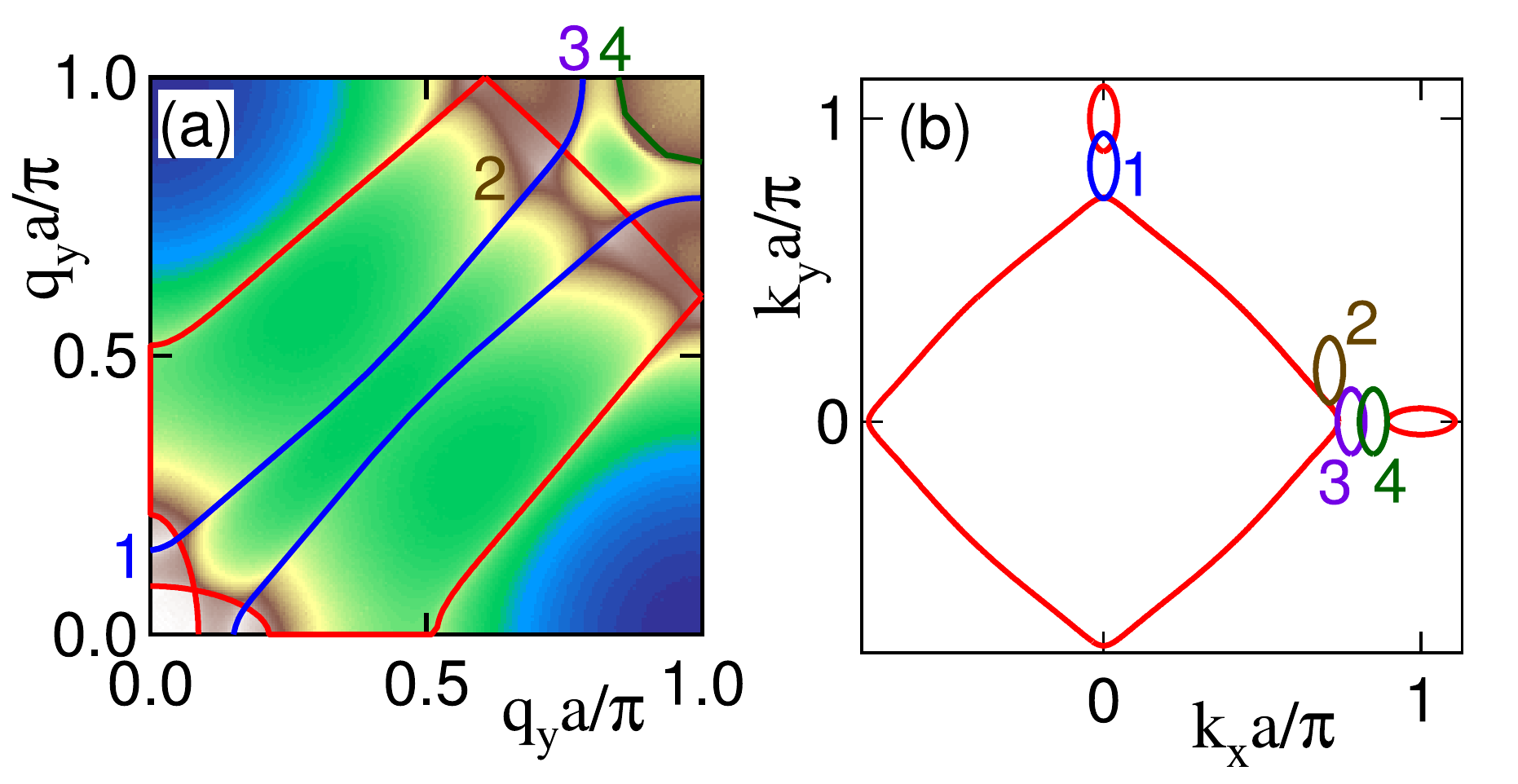}}
\vskip0.5cm
\caption{(Color online.)
(a) Susceptibility map for LSCO(1) (h) with nesting curves superposed. Red curves: 
${\bf q}=2{\bf k}_F$ nesting for both $\Gamma$-centered barrel and $(\pi ,0)$, $(0,\pi )$-centered 
pockets.  Blue lines = barrel-pocket nesting; green line = inter-pocket nesting.
(b) Corresponding Fermi surface map showing four ${\bf q}$-shifted pockets, labelled as in 
(a).  
}
\label{fig:5}
\end{figure}  

The complicated nesting map of the middle column near the VHS is explained in 
Fig.~\ref{fig:5}.  The dispersion is such as to produce an `extended VHS',\cite{EVH} 
which first intersects the $\Gamma\rightarrow (\pi ,0)$ axis at a point $(\pi -\delta ,0)$.  
For larger hole doping, the FS has two sheets, one a squarish barrel centered at $\Gamma$, 
the other a pocket centered at $(\pi ,0)$, Fig.~\ref{fig:5}(b).  The intrasheet ${\bf q}=2{\bf k}_F$ 
nesting maps are shown by the red curves in Fig.~\ref{fig:5}(a).  In addition to the one 
associated with the barrel FS, there are two overlapping segments associated with the 
pockets, but translated by the ${\bf q}$ folding to the vicinity of $\Gamma$.  As in the other 
frames of Figs.~2-4, these nesting curves exactly match some of the ridges seen in the 
susceptibility.  However, there are additional ridges, associated with nesting between 
two different FS segments.  Thus, the blue curves in Fig.~\ref{fig:5}(a) represent 
barrel-pocket nesting -- that is, the locus of the ${\bf q}$-vectors needed to shift the pocket 
until it is tangent to some point on the barrel FS.  For instance, the points labelled 
1,~2,~3 represent translations equivalent to those shown in Fig.~\ref{fig:5}(b).  
The new nesting curves can easily be found numerically, by requiring that the two FS 
sections have a common tangent at the point of osculation.  Similarly, the green curve near 
$(\pi ,\pi )$ represents inter-pocket nesting, with point 4 illustrating one particular 
nesting vector.  It can be seen that the full collection of nesting curves explains all of 
the ridge-like features seen in the susceptibility map, and in particular allow the 
determination of the points of maximal susceptibility, except for the above-noted 
commensurability effects.  Note however, that despite these complications, the 
susceptibility peak moves to $\Gamma$ in a finite doping range about the point where 
the antinodal electron pocket shrinks to zero width.

The above discussion can be summarized: in the doping between half filling and the 
VHS, a new susceptibility peak arises, associated with nesting of the antinodal parts 
of the large FS.  Two kinds of behavior are found: when the ANN is dominant, the peak 
evolves to $\Gamma$ at the VHS.  On the other hand, for 
LSCO(2) with small $t'$, the ANN peak is inherently weaker than the $(\pi ,\pi )$ peak, 
in which case the susceptibility remains commensurate at $(\pi ,\pi )$ from half filling 
almost to the VHS, then smoothly develops a $(\pi ,\pi -\delta )$ incommensurability.  
Yet even here, in the immediate vicinity of the VHS, the ANN peak at $\Gamma$ becomes 
dominant in a very limited doping range.  Hence, for most dispersions the susceptibility 
will have strong FM fluctuations near the VHS, while for other dispersions the 
fluctuations remain mostly AFM.

\section{Gutzwiller Calculation and $U_{eff}$}

In the GA+RPA calculation\cite{DLGS,QLif}, the electronic paramagnetic ground state energy 
is calculated in the GA\cite{Geb,KR} and then expanded to second order in the on-site and 
intersite magnetic fluctuations in the spirit of Vollhardt's Fermi liquid 
approach\cite{Voll}. Response functions are computed using linear
response in the presence of small external field.\cite{DLGS} One
obtains RPA like susceptibilities but with strong vertex corrections.
As shown in Appendix ~\ref{sec:spin-rotat-invar}
longitudinal and transverse susceptibilities are trivially related and
lead to the same Stoner criteria as
required by spin rotational invariance.  
In terms of a tensor bare transverse susceptibility (cf. 
Appendix~\ref{sec:spin-rotat-invar} and Appendix C)
\begin{eqnarray}
\chi_{0{\bf q}}&=&\Bigl(\matrix{\chi^0_{11}&\chi^0_{12}\cr
                             \chi^0_{21}&\chi^0_{22}\cr}\Bigr)
\nonumber \\
  &=&{1\over N}\sum_{{\bf k}}\Bigl(\matrix{1&E_{{\bf k},{\bf q}}\cr
                               E_{{\bf k},{\bf q}}&E^2_{{\bf k},{\bf 
q}}\cr} 
\Bigr){n_{{\bf k}+{\bf q}}-n_{{\bf k}}\over \epsilon_{{\bf k}+{\bf 
q}}-\epsilon_{{\bf k}}},
\label{eq:G3}
\end{eqnarray}
with $ E_{{\bf k},{\bf q}}= \epsilon^0_{{\bf k}+{\bf q}}+\epsilon^0_{{\bf 
k}}$, the GA+RPA dressed susceptibility $\chi_{\bf q}$ is given by 
\begin{equation}
\chi^{-1}_{\bf q}=\chi^{-1}_{0{\bf q}}- V^{+-}_{\bf q}.  
\label{eq:G4}
\end{equation}
Here the ratio of dressed to bare dispersion is given by $\epsilon_{{\bf 
k}}/\epsilon^0_{{\bf k}}=Z$, with the Gutzwiller renormalization factor
\begin{equation} \label{eq:z0}
Z=z_0^2=4\frac{\left[\sqrt{(x+D)(\frac{1-x}{2}-D)}+\sqrt{D(\frac{1-x}{2}-D)}\right]^2}{1-x^2}  
\end{equation}
which depends on  the GA double occupancy variational parameter $D$ and the doping $x$.
The interaction matrix is\cite{DLGS}
\begin{eqnarray}
V^{+-}_{\bf q}=\Bigl(\matrix{N_{\bf q}&M_{\bf q}\cr
                             M_{\bf q}&0\cr}\Bigr),
\label{eq:G5}
\end{eqnarray}
which is defined in Appendix~\ref{sec:spin-rotat-invar}.

While 
Eq.~\ref{eq:G4} is a tensor equation, it can be expanded into the form of Eq.~\ref{eq:G1} 
with $\chi_0=z_0^2\chi^0_{11}$ and $U_{eff}=U_{GA}$,
\begin{equation}
U_{GA}= (N_{\bf q}+2 M_{\bf q}\bar E_1+ M^2_{\bf q}(\bar E^2_2-\bar 
E^2_1)\chi_0/z_0^2)/z_0^2,
\label{eq:G6}
\end{equation}
with $\bar E_1=\chi^0_{12}/\chi^0_{11}$, $\bar
E^2_2=\chi^0_{22}/\chi^0_{11}$.  
The
$z_0$ factors appear now in $U_{GA}$ since we want to use 
the bare $\chi_0$ in Eq.~\ref{eq:G1}. All correlation effects are incorporated in
the definition of $U_{GA}$ which therefore can be viewed as a vertex corrected interaction
term in the magnetic $p-h$ channel.  Note that $U_{GA}$ is not strictly a pure 
interaction term, but is weighted by kinetic terms which enhance its ${\bf q}$-dependence.  

\begin{figure}
\leavevmode  
\resizebox{8.2cm}{!}{\includegraphics{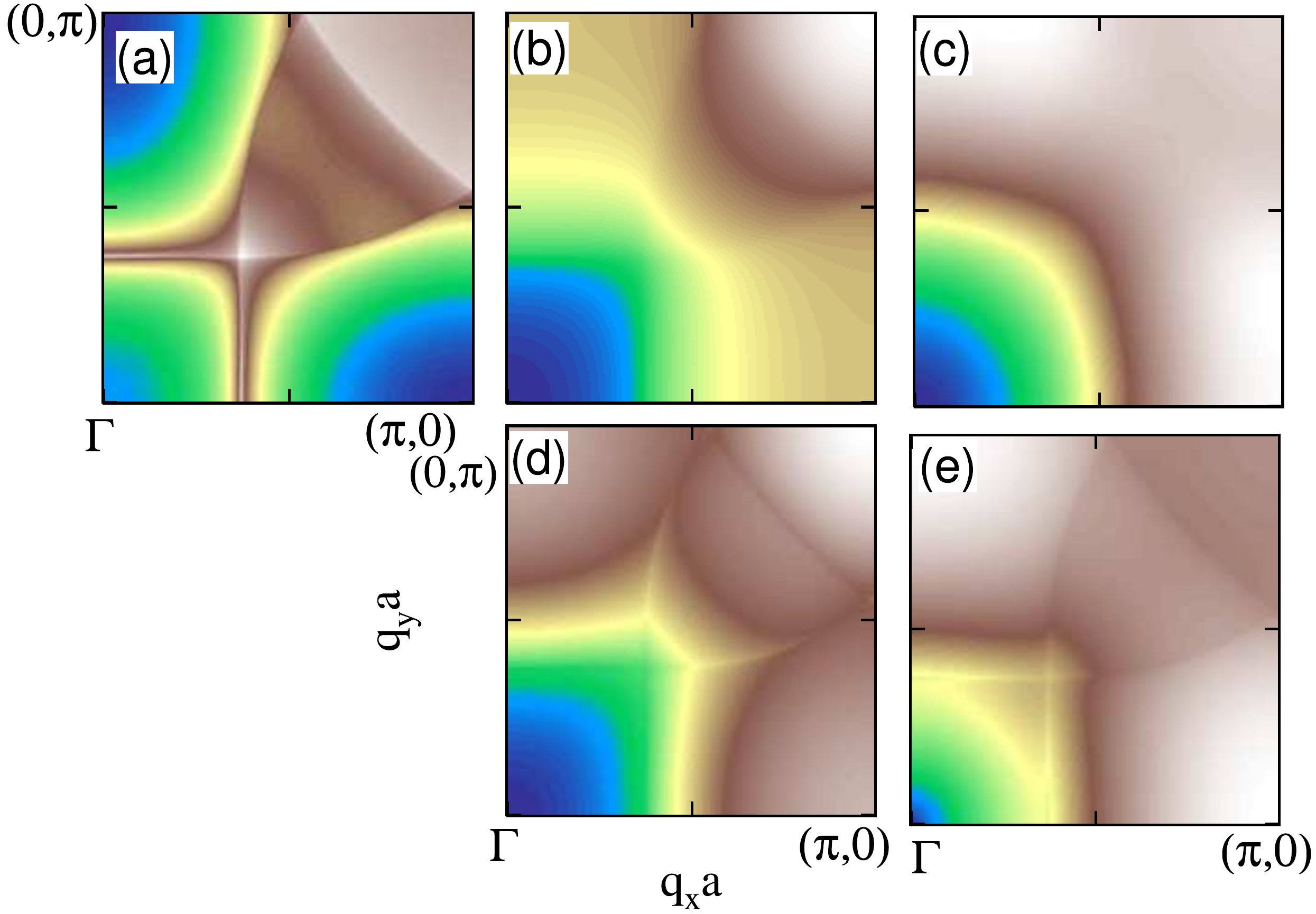}}
\vskip0.5cm
\caption{(Color online.)
Susceptibility maps (a) $\chi^0_{11}$, (b) $\chi^0_{12}$, and (c) $\chi^0_{22}$, along with 
derived quantities (d) $\bar E_1$ and (e) $\bar E_2$, as a function of ${\bf q}$ for Bi2201 
with $x=0.15$.
}
\label{fig:6}
\end{figure}  

\begin{figure}
\leavevmode  
\resizebox{8.2cm}{!}{\includegraphics{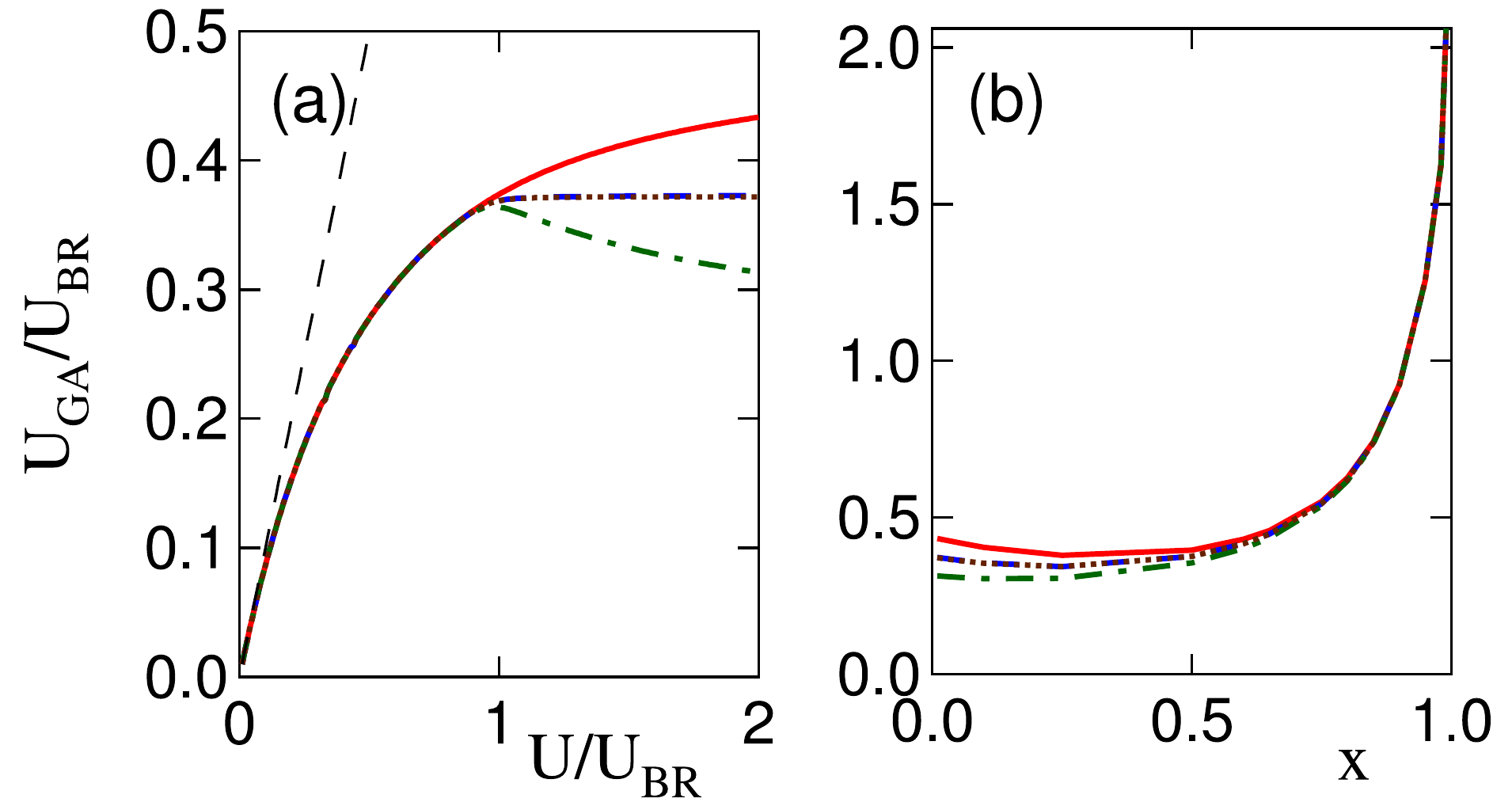}}
\vskip0.5cm
\caption{(Color online.)
(a) Variation of $U_{GA}$ with $U$ for $x=0.01$. (b) Variation of 
$U_{GA}$ with $x$ for $U=2U_{BR}$.  $U_{GA}$ varies with ${\bf q}$, 
and the figure shows ${\bf q}$ = $\Gamma$ (red solid line), $(\pi ,0)$ 
(blue dashed line), $(\pi /2,\pi /2)$ (brown dotted line), and $(\pi 
,\pi )$ (green dot-dashed line). The ${\bf q}$-points along the zone diagonal, $(\pi ,0)$ and $(\pi 
/2,\pi /2)$ have nearly identical behavior.
} \label{fig:7} \end{figure}

Figure~\ref{fig:6} compares $\chi^0_{11}$, $\chi^0_{12}$, $\chi^0_{22}$, and the 
derived $\bar E_1$, $\bar E_2$ -- the results are fairly insensitive to doping or band 
parameters.  The sharp structures associated with nesting show up only in $\chi^0_{11}$, 
while the other $\chi$'s are smooth, and the $\bar E$'s contain only a weak structure 
imposed from $\chi^0_{11}$.  Hence $U_{GA}$ remains a smooth, weakly varying function 
of ${\bf q}$, and the instabilities are controlled by the peaks in $\chi_0$.  
Figure~\ref{fig:7}(a) shows how $U_{GA}$ varies as the bare $U$ is increased.
At small $U$, $U_{GA}\rightarrow U$ and the conventional HF+RPA is recovered.
Again the results do not depend strongly on doping (Fig.~8(b)) or dispersion.  The general 
trend is that at large $U$ $U_{GA}$ tends to saturate, and the large-${\bf q}$ components of 
$U_{GA}$ are reduced most strongly, tending to favor instabilities nearer $\Gamma = (0,0)$, 
suggestive of ferromagnetic domains.  Fig.~\ref{fig:7} should be compared with 
Fig.~2 of Vilk and Tremblay\cite{VAT}, who find a similar saturation.  Since these 
authors employ a very different perscription for including vertex corrections, the 
similarity of our results gives considerable further confidence to the
trends we find. In addition since the Gutzwiller approximation has a
simple interpretation in terms of kinetic energy suppression due to
correlation our results shed further light on the physical meaning of
the vertex corrections.   

\section{Gutzwiller Magnetic Phase Diagrams of the Cuprates}

\begin{figure}
\leavevmode  
\resizebox{8.2cm}{!}{\includegraphics{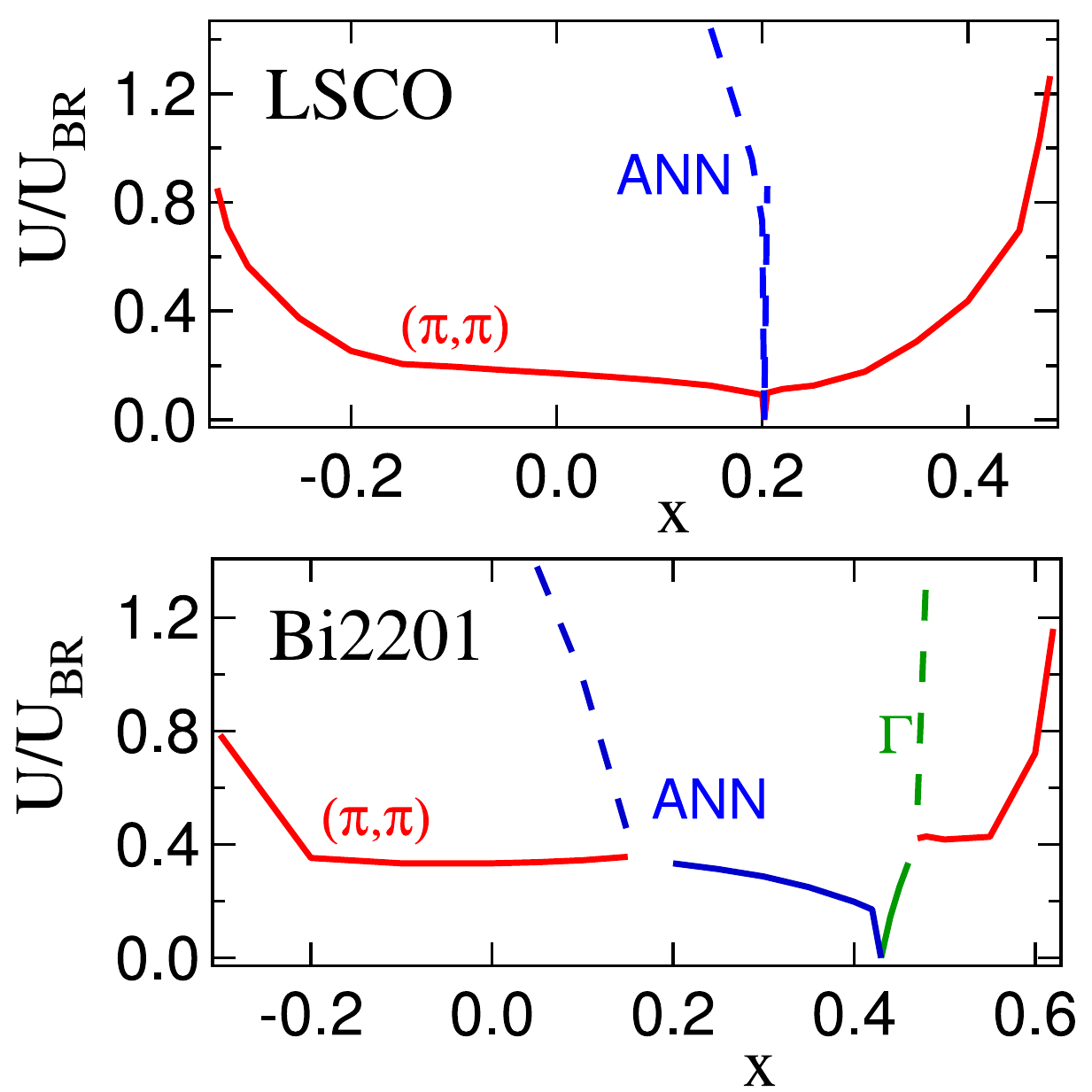}}
\vskip0.5cm
\caption{(Color online.)
Gutzwiller magnetic phase diagrams for (a) LSCO(2) and (b) Bi2201(2).  Dashed lines indicate 
metastable states -- extensions of the condition $U_{Gutz}\chi = 1$ for one phase beyond the 
point where another phase has become unstable.}
\label{fig:8}
\end{figure}  

Figure~\ref{fig:8} shows our main result, the magnetic phase diagrams of LSCO and 
Bi2201 calculated in the Gutzwiller GA+RPA approach, based on the Stoner criterion, Eq.~1.  
The phase diagram for Bi2201 is fairly generic, and we have found similar results for NCCO, 
Bi2212 [neglecting bilayer splitting] and SCOC, while the LSCO phase diagram is limited to 
the dispersion parameters of LSCO(1).  Thus, for electron doping the phase diagram is very 
simple, dominated by simple AF order with ${\bf q}$ very close to $(\pi ,\pi)$, while for hole 
doping there is a competition between the SAF and ANN order, except in LSCO, where SAF order 
dominates.  It is possible that ANN order is relevant to the `checkerboard' phase seen 
in STM studies of several cuprates, as discussed briefly below.

Results for HF+RPA are similar, but since $U_{eff}$ in Eq.~1 is not bounded ($U_{eff}=U$), 
there would be magnetic instabilities for all dopings, with transitions to the $(\pi ,0)$ 
and $\Gamma$ plateaus, as in Fig.~\ref{fig:1}.  In contrast, in the GA+RPA calculation $U_{GA}$ 
saturates, and magnetic order exists only in a limited doping range in the cuprates, 
primarily in the range where the $(\pi ,\pi )$ plateau exists.   However, this doping range 
encompasses the full doping range of relevance to cuprate physics -- including the 
overdoped regime -- if a doping independent $U\sim 6-8t$ is assumed.  Thus, to have a 
magnetic quantum critical point near optimal doping, one must go beyond the one-band 
GA+RPA approach.  The limited doping range for the magnetic phases has
also been seen in a $d=\infty$ study.\cite{QLif}
Analogously to HF+RPA, GA+RPA determines the transition line at which 
the mean field state becomes unstable, but the energies are evaluated at mean field level.  
Fluctuation contributions to the energy will most likely lower the energy of the 
paramagnetic (disordered) phase further reducing the stability range of the magnetic 
phases.

\section{Discussion}

There have by now been countless calculations concerning nanoscale phase separation or stripe phases 
in cuprates, and this is becoming a paradigm in many other correlated electronic systems as well.  We 
believe, however, that the present results are unique in providing a systematic phase diagram, 
covering the full doping dependence\cite{incomm} and all possible $q$-vectors for realistic band 
dispersions, and indeed finding that the dominant instabilities may be different in different 
cuprates.

\subsection{Extension to Charge Instabilities}

While the above has provided a thorough analysis of possible magnetic instabilities, 
nothing has been said about competing instabilities in the charge channel -- charge 
density waves (CDWs) or `stripes'.  Of course, one possibility is that charge order 
arises as a secondary effect following the spin order, with $\delta\rho\propto (\delta 
M)^2$.  The case where the charge instability is primary will be somewhat harder to 
analyze.  That is because the corresponding Stoner factor 
analysis is likely to require an extension beyond the Hubbard model, and hence involve 
greater uncertainty in the choice of $U_{eff}({\bf q})$.  For charge instabilities, the 
corresponding Stoner denominator is nothing but the (zero frequency) dielectric constant 
$\epsilon /\epsilon_0=1+V({\bf q})\chi_0({\bf q},\omega =0)$, where $\epsilon_0$ is a background 
dielectric constant and $V({\bf q})$ is a Coulomb potential.  When $V({\bf q})$ is taken as either the 
Hubbard $U$ or a long-range Coulomb interaction,\cite{plas} $\epsilon$ is always positive and 
there is no instability.

Moreover, from the theory of dielectric stability, it is known that $\epsilon$ cannot 
fall in the range between 0 and 1 [equivalently $\epsilon^{-1}\le 1$], and an 
instability $\epsilon=0$ must be approached through negative values of 
$\epsilon$.\cite{DKM}  While a purely electronic instability could still arise via 
inclusion of local field effects, the most natural situation arises when electron-phonon 
coupling is included. A simple $s$ wave instability will be suppressed
by the on-site Coulomb repulsion as shown in Ref.~\onlinecite{DLGS}
but one can still have instabilities in other channels. To the extent
that the matrix elements which determine the channel have a smooth
behavior one will have a Stoner denominator similar to
Eq.~(\ref{eq:G1}) with $U^{(ep)}_{eff}\sim 
g^2D_0/\epsilon_e\Omega_{ph}^2$, where $g$ is the electron-phonon coupling parameter, 
$D_0$ the corresponding bare phonon propagator, $\Omega_{ph}$ a bare phonon 
frequency, and $\epsilon_e$ an electron-electron dielectric constant.\cite{DLGS} 

Thus, since the instability is also controlled by a Stoner factor, the results of the 
present paper will also apply for charge instabilities.  This is consistent with the 
common expectation that CDWs and spin density waves (SDWs) are controlled by the same 
nesting instabilities. Stated differently, the Stoner criterion is a formal expression 
of the idea that $P/K\ge 1$ for an instability, where $P$, the interaction energy, is 
represented by $U_{eff}$ while the kinetic energy $K$ is measured via $\chi_0^{-1}$.  The 
kinetic energy is closely tied to the band structure, and carries important 
material-specific information.  In contrast, the potential energy is
fairly featureless, leading to smoothly varying $U_{eff}({\bf q})$.  Thus, the locations of the 
instabilities are controlled by peaks in $\chi_0$, and a ${\bf q}$-dependent $U$ can only 
shift the dominant instability between two peaks of comparable height. In the remainder 
of this section, when we discuss comparison to experiment, it will be seen that the 
present model has correctly determined the dominant ${\bf q}$ values, but that in several 
cases experiment points more toward CDWs than SDWs. 

\subsection{Stripes vs Checkerboards}

There have been a number of recent hints that the `stripe' order in underdoped 
La$_{2-x}$A$_x$CuO$_{4+\delta}$, A = Sr (LSCO) or Ba (LBCO) is not the same phenomenon as 
the `checkerboard' order found in Sr$_2$CuO$_2$Cl$_2$ (SCOC) and 
Bi$_2$Sr$_2$CaCu$_2$O$_8$ (Bi-2212).  Thus, resonant soft x-ray scattering   
experiments find evidence of charge order in LBCO (stripes)\cite{Abbayes} but not in 
Ca$_2$CuO$_2$Cl$_2$ (checkerboards)\cite{Abbano}, while evidence for time reversal symmetry 
breaking has been found in YBa$_2$Cu$_3$O$_{7- \delta}$ (YBCO), but not in LSCO\cite{MUWL}.  
The present results suggest a connection between the ANN and checkerboard phases, in that 
the periodicity of the latter also scales with the antinodal nesting vector\cite{Kyle},
and a further connection between the $(\pi ,\pi )$ plateau and stripes.  
Thus it is quite interesting to observe that the ANN phase is virtually absent for the LSCO(2) 
dispersion, where evidence for conventional stripes is strongest.  

\subsection{Conventional and VHS nesting}

In 1D systems the susceptibility diverges at ${\bf q}=2{\bf k}_F$ at all dopings, due to perfect 
(flat band) nesting.  The present results generalize this to 2D systems: the dominant 
instability is generally associated with double nesting, but since $\chi_0$ in general 
does not diverge, a finite coupling $U$ is required to drive an instability.  An 
exception is VHS nesting\cite{RiSc}, for which $\chi_0$ has a logarithmic divergence.  

At the VHS, there are two competing instabilities, AF at $(\pi ,\pi )$ and FM at 
$\Gamma$, corresponding to the magnetic branch of the $SO(8)$ phase diagram of the 
cuprates, and these are responsible for the commensurate pinning near $(\pi ,\pi )$ and 
$\Gamma$, respectively.\cite{SO8}  Doped away from the VHS, these instabilities evolve 
into the two dominant peaks of the susceptibility, and can be considered as `generalized 
VHS nesting'. As such, they dominate the magnetic physics of 
the cuprates over the full doping range, coextensive with the limits of the $(\pi ,\pi 
)$-plateau in the susceptibility.  

\subsection{VHS and FM}

Recently, Storey {\it et al.}\cite{TallStorey} proposed that the generic behavior of 
high-T$_c$ cuprates could be understood if the pairing interaction (or pairing energy 
cutoff) falls off rapidly near the VHS.  A recent experiment\cite{PaKa} does indeed 
confirm that $x_{opt}$ scales with $T_c$, but $x_{opt}/x_{VHS}<1$. The present results 
suggest that most cuprates will have strong FM fluctuations near the VHS, which are 
incompatible with simple d-wave superconductivity.  Empirical evidence for FM fluctuations 
has been noted previously\cite{FMcomp}.  While some previous calculations\cite{FMV} have 
found evidence of ferromagnetism near a VHS, others\cite{IKI} have suggested that FM 
instabilities were unlikely in competition with incommensurate susceptibility.  The present 
calculations confirm that a dominant FM susceptibility should be present near the VHS.

\subsection{Limitations of Present Approach}

\subsubsection{$k_z$-dispersion}

The advantage of two-dimensional materials is that it is straightforward to 
display the susceptibility maps and nesting curves.  In 3D, the curves become {\it 
nesting surfaces} and the susceptibility maps are 4-dimensional.  One can speculate that the 
dominant susceptibility peaks correspond to triple-nesting points.  For quasi-2D 
materials, it should be possible to analyze a series of cuts perpendicular to the 
(weakly-dispersing) $z$-axis.  

\subsubsection{Nanoscale phase separation}

In principle, the present results could provide information on nanoscale phase 
separation (NPS) as well.  One model of NPS is that there are different 
instabilities associated with particular dopings [e.g., half filling and the VHS], 
and that these two phases are more stable than uniform phases at intermediate 
dopings.  In this case, the two end phases could still be described by nesting 
maps, only at particular dopings.  For instance, in the HF+RPA analysis of LSCO(2), the FM 
phase is stable only very close to the VHS, and could lead to NPS with a second phase 
at the undoped insulator.  

\subsubsection{Away from the Instability Threshold -- Towards Strong Correlations}

The Stoner criterion determines which ${\bf q}$ value is most unstable, and the minimum $U$ 
needed to drive that instability.  However, as $U$ increases above threshold, the ${\bf q}$ of 
the ordered phase may shift.  Thus, at half filling, when a full gap can be 
formed, ${\bf q}$ will no longer be determined by best nesting, but by the largest gap.  This 
tends to favor more commensurate ${\bf q}$ values, leading to a pinning of ${\bf q}$ at these 
commensurate values over a wide range of parameters.  We find that as $U$ increases, there is
a first order transition from an incommensurate phase with Fermi surface pockets to a
commensurate [$(\pi ,\pi )$ or $(\pi ,0)$] phase that is fully gapped.\cite{RM10} Away from
half filling, Luttinger's theorem ensures the persistence of a Fermi surface, so nesting
instabilities should persist over a wider range of $U$'s.

Furthermore, as we have seen above, the ${\bf q}$-value corresponding to the largest 
$U_{eff}({\bf q})\chi_0({\bf q})$ can shift with $U$.  Typically, for $U_{GA}$ the shift is to a 
smaller ${\bf q}$-value, associated with an instability towards FM order (in the magnetic 
channel) or NPS (in the charge channel).

\section{Conclusion}

The present results provide a constructive scheme for identifying the dominant nesting 
instabilities for any two dimensional material.  Clearly, for Fermi surfaces with 
multiple sections, a large number of nesting curves are possible, leading to extremely 
complicated susceptibility maps.  Nevertheless, the present scheme will automatically 
sort out the possible double nesting peaks and follow their evolution with doping.  This 
should allow a much more detailed understanding of 2D phase diagrams, particularly for 
magnetic phases, where the interaction $U$ has negligible ${\bf q}$-dependence.

The good agreement of the GA+RPA calculations with more detailed variational calculations 
at half filling\cite{RM10} and with exact and numerical results in $d=\infty$,\cite{QLif} 
and of the AF+SC model in electron doped cuprates with experiments gives us confidence in 
the model.  Accordingly, we note the following three points.  First, within GA+RPA, the 
paramagnetic state is unstable over the full hole-doping range in the cuprates, including 
overdoped.  To avoid this conclusion, and restore a Fermi liquid phase in the overdoped 
regime, it may be necessary to include non Gaussian fluctuations\cite{ACCG} or an additional doping dependence for 
$U$.  The origin of any such doping dependence lies outside the GA treatment of the 
Hubbard model.  Second, the $(\pi ,\pi)$ phase, or its incommensurate extension, is 
unstable against a competing ANN order in [most] hole-doped 
cuprates.  Third, there is a material dependence to the phase diagram, and LSCO may have a 
very different doping dependence than other cuprates.  

The results have been applied to a number of model dispersions for the cuprates.  A 
possible pseudogap candidate has been identified and a distinction between stripes and 
checkerboards proposed.  These findings will be discussed in greater detail in ensuing 
publications.

This work is supported by the US Department of Energy, Office of
Science, Basic Energy Sciences contract DE-FG02-07ER46352, and
benefited from the allocation of supercomputer time at NERSC,
Northeastern University's Advanced Scientific Computation Center
(ASCC).  RSM's work has been partially funded by the Marie Curie
Grant PIIF-GA-2008-220790 SOQCS, while GS' work is supported by
the Vigoni Program 2007-2008 of the Ateneo Italo-Tedesco
Deutsch-Italienisches Hochschulzentrum.

\appendix
\section{Notes to Table I: Band Parameters}
\label{sec:band-parameters}

\begin{table}
\caption{I. Band Parameter Sets}
\begin{tabular}{||c|c|c|c|c|c||}
\hline
Parameter& NCCO & Bi-2201(1) & Bi-2201(2)  &  LSCO(1) & LSCO(2) 
   \\      \hline          %
t        &    420    &  250    &  435       & 195.6     &  419.5 \\  \hline  
t'       &   -100    &  -55    & -120       & -18.5     &  -37.5    \\  \hline
t''      &     65    &   27.5  &   40       &  15.7     &   18      \\  \hline
t'''     &      7.5  &    0    &    0       &  17.5     &   34      \\  \hline
t''''    &      0    &    0    &    0       &   4.35    &    0      \\  \hline
Z    &  0.5  & 1 & 0.5 &1 &0.5  \\   \hline
Ref.: & \onlinecite{Arun3} & \onlinecite{ZX01} & \onlinecite{Tanmoy1}& 
\onlinecite{Norm2}&\onlinecite{Arun3} 
   \\      \hline          %
\hline
\end{tabular}  
\end{table} 
In Table I, all hopping parameters are given in meV.  The $t''''$ term for 
LSCO(1) is the coefficient of a term $c_x(2a)c_y(2a)$ in Eq.~\ref{eq:2}.
NCCO, LSCO(2), and Bi2201(2) data sets are fit to LDA band dispersions of the 
near-Fermi level antibonding CuO$_2$ band, and are appropriate for the Gutzwiller 
analysis.  The data sets LSCO(1) and Bi2201(1) are taken from fits to 
experimental ARPES data.  To the extent that the bands are renormalized by a 
${\bf q}$-independent factor $Z$, the effective bare susceptibilities $Z\chi$ 
can be used in the Gutzwiller analysis, while $\chi_{0M}$ occurs at the same 
${\bf q}$.  

Recently it has been suggested that the enhanced 
nesting, such as observed in LSCO(1), may be associated with stripe 
formation\cite{ZKI}.

\section{Gutzwiller plus random-phase approximation formalism.}
\label{sec:spin-rotat-invar}

Here we sketch the GA+RPA formalism and  define the relevant
quantities. For a review see Ref.~\onlinecite{jose06}. 

 The
susceptibility can be computed in the longitudinal\cite{DLGS} or in the
transverse channel.~\cite{sei04} As shown bellow  both results are
equivalent as dictated by spin rotational invariance on a paramagnetic
(singlet) ground state.

For later use we define the density matrix associated with the unprojected Slater
determinant, $|\phi \rangle$, as  $$\rho_{ij}^{\sigma\sigma'}\equiv
\langle\phi| c_{j\sigma'}^\dagger c_{i\sigma} | \phi\rangle $$ and  as a shorthand 
we define $\rho_{ij\sigma}\equiv\rho_{ij}^{\sigma\sigma}$.

We start from the spin-rotational invariant 
Gutzwiller energy functional for the Hubbard model defined in terms of
$\rho_{ij}^{\sigma\sigma'} $ and the double occupancy in the Gutzwiller
variational state $D_i$. As
derived e.g. in Ref.~\onlinecite{sei04} the functional reads,
\begin{displaymath}
E^{GA}= \sum_{i,j}
t_{ij} \langle\phi|{\bf \Psi_i}^\dagger {\bf z_{i}}
{\bf z_{j}}{\bf \Psi_j}|\phi\rangle + U\sum_{i} D_i  .
\end{displaymath}
where we have defined the spinor operators
\begin{displaymath}
{\bf \Psi_i}^\dagger = (c_{i\uparrow}^\dagger , c_{i\downarrow}^\dagger)
\,\,\,\,\,\,  {\bf \Psi_i} = \left(\begin{array}{c} c_{i\uparrow} \\
c_{i\downarrow} \end{array}\right)
\end{displaymath}
and  the ${\bf z}$-matrix 
\begin{widetext}
\begin{eqnarray*}
{\bf z}_i&=& \left( \begin{array}{cc}
z_{i\uparrow}\cos^2\frac{\varphi_i}{2}+z_{i\downarrow}
\sin^2\frac{\varphi_i}{2} &
\frac{S_i^-}{2S_i^z}[z_{i\uparrow}-z_{i\downarrow}]\cos\varphi_i \\
\frac{S_i^+}{2S_i^z}[z_{i\uparrow}-z_{i\downarrow}]\cos\varphi_i&
z_{i\uparrow}\sin^2\frac{\varphi_i}{2}+z_{i\downarrow}\cos^2\frac{\varphi_i}{2}
\end{array} \right) \\
\tan^2\varphi_i&=&\frac{S_i^+S_i^-}{(S_i^z)^2}.
\end{eqnarray*}
with the $z$ factors given by
\begin{displaymath}
z_{i\sigma} = \frac{\sqrt{(1-\rho_{ii}+D_i)(\frac{1}{2}\rho_{ii}+\frac{S_i^z}{\cos(\varphi_i)}-D_i)}+\sqrt{D_i(\frac{1}{2}\rho_{ii}-\frac{S_i^z}{\cos(\varphi_i)}-D_i)}}
{\sqrt{(\frac{1}{2}\rho_{ii}+\frac{S_i^z}{\cos(\varphi_i)})(1-\frac{1}{2}\rho_{ii}-
\frac{S_i^z}{\cos(\varphi_i)})}}
\end{displaymath}
\end{widetext}
and for clarity spin expectation values are denoted by
$S_i^+=\rho_{ii}^{\uparrow,\downarrow}$,
$S_i^-=\rho_{ii}^{\downarrow,\uparrow}$,
$S_i^z=(\rho_{ii}^{\uparrow,\uparrow}-\rho_{ii}^{\downarrow,\downarrow})/2$,
and
$\rho_{ii}=\rho_{ii}^{\uparrow,\uparrow}+\rho_{ii}^{\downarrow,\downarrow}$.

In the limit of a vanishing rotation angle $\varphi$ the ${\bf z}$-matrix
becomes diagonal and the renormalization factors
reduce to those of the standard Gutzwiller approximation.

For a homogeneous, paramagnetic system ($z_{i\sigma}\equiv z_0$, $\varphi_i=0$) 
the expansion of the Gutzwiller energy 
functional up to second order in the particle-hole excitations 
reads, 
\begin{equation}
\label{eq:ega2}
\delta E^{GA}=\delta E^{ff}+ 
 \delta E^{GA}_c + \delta E^{GA}_{lo} + \delta E^{GA}_{tr}
\end{equation}
The first contribution is the free fermion part,
$$
\delta E^{ff}= \sum_{k>k_F , k'<k_F ; \sigma,\sigma'}(\epsilon_{\kvec}-\epsilon_{\kvec'})\delta 
\rho_{\kvec\kvec'}^{\sigma\sigma'}\delta \rho_{\kvec'\kvec}^{\sigma'\sigma}.
$$
$\epsilon_{\kvec}$ denotes the dispersion
of the Gutzwiller quasiparticles whereas $\epsilon^0_{\kvec}$ corresponds
to the unrenormalized dispersions, i.e. $\epsilon_{\kvec} = z_0^2
\epsilon^0_{\kvec}$.

The remaining part in Eq.~(\ref{eq:ega2}) is due to quasiparticle interactions and  
separates into contributions from the charge ($c$),  the 
longitudinal ($lo$) and transverse ($tr$) spin channel.
$\delta E^{GA}_c$ and $\delta E^{GA}_{lo}$ have been derived in detail in 
Ref.~\onlinecite{DLGS}.

 For our purposes we report in the following the
expansion in the longitudinal channel:
\begin{eqnarray}
\delta E_{lo}^{GA} &=& \frac{1}{2N} \sum_{\qvec}
\left(\begin{array}{c}
\delta S^{z}_\qvec \\ \delta T^{z}_{\qvec}
\end{array}\right)
\underline{\underline{V_q^{zz}}}
\left(\begin{array}{c}
\delta S^{z}_{-\qvec} \\ \delta T^{z}_{-\qvec}
\end{array}\right)
\label{ES} \\
\underline{\underline{V_q^{zz}}}&=&
\left(\begin{array}{cc}
2 N_\qvec &  2 z_0 z'_t \\
2 z_0 z'_t & 0
\end{array}\right) \label{eq:vzz}
\end{eqnarray}
and in the transverse magnetic channel which has been derived in 
Ref. \onlinecite{sei04}
\begin{eqnarray}
\delta E_{tr}^{GA} &=& \frac{1}{N} \sum_{\qvec}
\left(\begin{array}{c}
\delta S^{+}_\qvec \\ \delta T^{+}_{\qvec}
\end{array}\right)
\underline{\underline{V_q^{+-}}}
\left(\begin{array}{c}
\delta S^{-}_{-\qvec} \\ \delta T^{-}_{-\qvec}
\end{array}\right)
\label{eskspace}\\
\underline{\underline{V_q^{+-}}} &=&
\left(\begin{array}{cc}
N_\qvec &  z_0 z'_t \\
z_0 z'_t & 0 
\end{array}\right) .\label{eq:vpm}
\end{eqnarray}
The relevant fluctuations have been defined as 
\begin{eqnarray*}
\delta S^z_q &=& \frac{1}{2}\sum_{k\sigma}\sigma \delta 
\rho_{k+q,k}^{\sigma,\sigma}, \\
\delta T^z_q &=& \frac{1}{2}\sum_{k\sigma} \sigma\left(\varepsilon_{k + q}^0+
\varepsilon_k^0\right)
\delta \rho_{k+q,k}^{\sigma,\sigma}\\
\delta S^\sigma_q &=& \sum_k \delta \rho_{k+q,k}^{\sigma,-\sigma}, \\
\delta T^\sigma_q &=& \sum_k \left(\varepsilon_{k + \sigma q}^0+\varepsilon_k^0\right)
\delta \rho_{k+q,k}^{\sigma,-\sigma}.
\end{eqnarray*}

The 11 element of the interaction kernel is given by
\begin{equation}
N_{\bf q}=(z'_t)^2 N_{1{\bf q}}+2z_0z''_t e_0
\label{eq:G15}
\end{equation}
with
\begin{equation}
e_0\equiv{1\over N}\sum_{{\bf k}\sigma}\epsilon^0_{{\bf k}\sigma}n_{{\bf k}\sigma}=U_{BR}/8.
\label{eq:G16}
\end{equation}
Here $U_{BR}$ is the Brinkman-Rice energy.  $N_1$ is similar, but requires separate 
averages for different components of the energy:
\begin{eqnarray}
N_{1{\bf q}}&=&{1\over N}\sum_{{\bf k}\sigma}\epsilon^0_{{\bf k+q}\sigma}n_{{\bf k}\sigma}
 \\
&=&-t[c_{qx}+c_{qy}]\bigl<{c_{kx}+c_{ky}}\bigr>
-4t'c_{qx}c_{qy}\bigl< c_{kx}c_{ky}\bigr>
\nonumber \\
&-&t''[c_{2qx}+c_{2qy}]\bigl<{c_{2kx}+c_{2ky}}\bigr>
\nonumber \\
&-&2t'''[c_{qx}c_{2qy}+c_{qy}c_{2qx}]\bigl<{c_{kx}c_{2ky}+c_{ky}
c_{2kx}}\bigr>,\nonumber
\label{eq:G17}
\end{eqnarray}
in an obvious notation.

Furtheron we define the susceptibility matrices for the bare 
time-ordered correlation functions both in the longitudinal
\begin{displaymath}
{\bf \chi}_q^{0,lo}(t)=\frac{-i}{N} \left( \begin{array}{cc}
\langle {\cal T} \hat S^z_q(t) \hat S_{-q}^z(0)\rangle_0 & \langle
{\cal T}\hat S^z_q(t) \hat T_{-q}^z(0)
 \rangle_0 \\
\langle {\cal T} \hat T^z_q(t) \hat S_{-q}^z(0)\rangle_0 & \langle
{\cal T} \hat T^z_q(t) \hat T_{-q}^z(0)
 \rangle_0
\end{array} \right),
\end{displaymath}
and in the transverse channel
\begin{displaymath}
{\bf \chi}_q^{0,tr}(t)=\frac{-i}{N} \left( \begin{array}{cc}
\langle {\cal T} \hat S^+_q(t) \hat S_{-q}^-(0)\rangle_0 & \langle
{\cal T} \hat S^+_q(t) \hat T_{-q}^-(0)
 \rangle_0 \\
\langle {\cal T} \hat T^+_q(t) \hat S_{-q}^-(0)\rangle_0 & \langle
{\cal T} \hat T^+_q(t) \hat T_{-q}^-(0)
 \rangle_0
\end{array} \right),
\end{displaymath}
where a hat has been added to distinguish fluctuations 
($\delta S^+_q$) form operators  ($\hat S^+_q$).

The longitudinal  
susceptibility describes the $\Delta m_z=0$, singlet to triplet
excitations of the  paramagnetic state while the transverse describes   
$\Delta m_z=\pm 1$ spin
excitations.  Spin rotational 
invariance dictates that these excitations should be degenerate.

One obtains for  these correlation functions
\begin{displaymath}
\label{suslo}
\chi^{0,lo}_{\qvec}=  \frac{1}{4N} \sum_{\kvec\sigma}\left(\begin{array}{cc}
1 & \epsilon^0_{\kvec} + \epsilon^0_{\kvec+\qvec}\\
\epsilon^0_{\kvec} + \epsilon^0_{\kvec+\qvec} & (\epsilon^0_{\kvec} 
+ \epsilon^0_{\kvec+\qvec})^2 \end{array}\right)
 \frac{n_{\kvec+\qvec,\sigma} - n_{\kvec\sigma}}
{\omega+\epsilon_{\kvec+\qvec}-\epsilon_{\kvec}}
\end{displaymath}
and
\begin{displaymath}
\label{sustr}
\chi^{0,tr}_{\qvec}=  \frac{1}{N} \sum_{\kvec}\left(\begin{array}{cc}
1 & \epsilon^0_{\kvec} + \epsilon^0_{\kvec+\qvec}\\
\epsilon^0_{\kvec} + \epsilon^0_{\kvec+\qvec} & (\epsilon^0_{\kvec\sigma} 
+ \epsilon^0_{\kvec+\qvec})^2 \end{array}\right)
 \frac{n_{\kvec+\qvec,\uparrow} - n_{\kvec, \downarrow}}
{\omega+ \epsilon_{\kvec+\qvec}-\epsilon_{\kvec}}.
\end{displaymath}

For the non-interacting Gutzwiller quasiparticles spin-rotational invariance
is thus guaranteed from the relation
\begin{equation}
\chi^{0,lo}_{\qvec} = \frac{1}{2} \chi^{0,tr}_{\qvec}
\end{equation}

This identity is preserved within the GA+RPA when we compute the interacting
susceptibilities from the RPA series
\begin{eqnarray*}
\chi_{\qvec}^{lo} = \chi_\qvec^{0,lo} -
\chi_\qvec^{0,lo} V^{zz}_\qvec \chi^{lo}_\qvec \\
\chi_{\qvec}^{tr} = \chi_\qvec^{0,tr} -
\chi_\qvec^{0,tr} V^{+-}_\qvec \chi^{tr}_\qvec
\end{eqnarray*} 
since from Eqs. (\ref{eq:vzz}, \ref{eq:vpm})  the
interaction kernels are  
related by $V^{zz}=2 V^{tr}$. Clearly the energies of particle hole excitations
described by these equations are degenerate. In particular they lead to
the same Stoner criteria in both channels.

\section{Derivatives of the Gutzwiller approximation $z$ factors}
\label{sec:parameters-u_ga}

The expressions involve the following derivatives of $z_{i\sigma}$ in
the longitudinal channel   
\begin{equation}
z'\equiv{\partial z_{i\sigma}\over\partial\rho_{ii\sigma}},
\label{eq:G10}
\end{equation}
\begin{equation}
z'_{+-}\equiv{\partial z_{i\sigma}\over\partial\rho_{ii-\sigma}},
\label{eq:G11}
\end{equation}
\begin{equation}
z''_{++}\equiv{\partial^2 z_{i\sigma}\over\partial\rho^2_{ii\sigma}},
\label{eq:G12}
\end{equation}
\begin{equation}
z''_{+-}\equiv{\partial^2 z_{i\sigma}\over\partial\rho_{ii\sigma}
\partial\rho_{ii-\sigma}},
\label{eq:G13}
\end{equation}
\begin{equation}
z''_{--}\equiv{\partial^2 z_{i\sigma}\over\partial\rho^2_{ii-\sigma}}.
\label{eq:G14}
\end{equation}
The derivatives in the transverse channel can be related to those in
the longitudinal channel as follows,
\begin{eqnarray*}
 z'_t &\equiv &
\frac{\partial z_{i,\sigma,-\sigma}}{\partial \rho_{ii}^{-\sigma,\sigma}}
=z'-z'_{+-}= \frac{2\delta}{1-\delta^2}\left(\frac{1}{z_0} - z_0\right), \\
z''_t &\equiv & \frac{\partial^2 z_{i,\sigma \sigma}}
{\partial \rho_{ii}^{\sigma, -\sigma}\partial \rho_{ii}^{-\sigma, \sigma}}
=\frac12(z_{++}''+ 2z_{+-}''+z_{--}'')\\
&=&
\frac{2z_0}{(1-\delta^2)^2}\left\lbrace 1-2\delta^2\left(\frac{1}
{z_0^2} - 1\right)\right\rbrace \\
&&-\frac{1}{2}\frac{z_0}{(1-\delta-2D)^2}.
\end{eqnarray*}
On the right we have given explicit expressions in terms of  
$\delta$, the doping measured with respect to half-filling and the
double occupancy $D$ in the paramagnetic state.

\end{document}